\documentclass[twocolumn,citeautoscript,prl,superscriptaddress,amsmath,amssymb,8pt]{revtex4-1}

\usepackage{graphicx}
\usepackage{color}
\usepackage[dvipsnames]{xcolor}
\usepackage{upgreek}
\usepackage{float}
\usepackage{upgreek}
\usepackage{amsmath,amssymb}
\newcommand{\C}[1]{\ensuremath{^{#1}}C}
\newcommand{\degC}[0]{\ensuremath{^{\circ}}C}
\newcommand{\ket}[1]{\ensuremath{\left| #1 \right\rangle}}
\begin{document}

\title{High precision single qubit tuning via thermo-magnetic field control}

\author{David A. Broadway}
\affiliation{Centre for Quantum Computation and Communication Technology, School of Physics, University of Melbourne, Parkville, VIC 3010, Australia}
\affiliation{School of Physics, University of Melbourne, Parkville, VIC 3010, Australia}	

\author{Scott E. Lillie}
\affiliation{Centre for Quantum Computation and Communication Technology, School of Physics, University of Melbourne, Parkville, VIC 3010, Australia}
\affiliation{School of Physics, University of Melbourne, Parkville, VIC 3010, Australia}	

\author{Nikolai Dontschuk}
\affiliation{Centre for Quantum Computation and Communication Technology, School of Physics, University of Melbourne, Parkville, VIC 3010, Australia}
\affiliation{School of Physics, University of Melbourne, Parkville, VIC 3010, Australia}

\author{Alastair Stacey}
\affiliation{Centre for Quantum Computation and Communication Technology, School of Physics, University of Melbourne, Parkville, VIC 3010, Australia}
\affiliation{Melbourne Centre for Nanofabrication, Clayton, VIC 3168, Australia}

\author{Liam T. Hall}
\affiliation{School of Physics, University of Melbourne, Parkville, VIC 3010, Australia}

\author{Jean-Philippe Tetienne}
\email{jtetienne@unimelb.edu.au}
\affiliation{School of Physics, University of Melbourne, Parkville, VIC 3010, Australia}

\author{Lloyd C. L. Hollenberg}
\affiliation{Centre for Quantum Computation and Communication Technology, School of Physics, University of Melbourne, Parkville, VIC 3010, Australia}
\affiliation{School of Physics, University of Melbourne, Parkville, VIC 3010, Australia}	

\begin{abstract}	
	Precise control of the resonant frequency of a spin qubit is of fundamental importance to quantum sensing protocols. We demonstrate a control technique on a  single nitrogen-vacancy (NV) centre in diamond where the applied magnetic field is modified by fine-tuning a permanent magnet's magnetisation via temperature control. Through this control mechanism, nanoscale cross-relaxation spectroscopy of both electron and nuclear spins in the vicinity of the NV centre are performed. We then show that through maintaining the magnet at a constant temperature an order of magnitude improvement in the stability of the NV qubit frequency can be achieved. This improved stability is tested in the polarisation of a small ensemble of nearby \C{13} spins via resonant cross-relaxation and the lifetime of this polarisation explored. The effectiveness and relative simplicity of this technique may find use in the realisation of portable spectroscopy and/or hyperpolarisation systems.
\end{abstract}
\maketitle

Progress in quantum information processing (QIP) and quantum sensing (QS) has driven interest in spin qubits\,\cite{Morello2010,Hill2015,Cole2009,Bienfait2015}, which are attractive candidates to both of these endeavours. One such qubit is the nitrogen-vacancy (NV) centre in diamond\,\cite{Doherty2013}, which has long coherence times at room temperature, and has demonstrated the capability of measuring a range of nanoscale field values from magnetic fields\,\cite{Schirhagl2014,Rondin2014,Maze2008,Balasubramanian2008}, electric fields\,\cite{Dolde2011}, temperature\,\cite{Neumann2013,Toyli2013,Kucsko2013} and can even achieve sub-mHz spectral resolution in nuclear magnetic resonance\,\cite{Schmitt2017,Boss2017,Bucher2017}. Application of the NV centre in QIP and QS requires control of the spin sub-levels of the system via the Zeeman effect\,\cite{Balasubramanian2008}. To implement this control, a permanent magnet is typically used as it is simpler and more cost-effective than an electromagnet (for small fields $B < 0.1$T), however, this poses two problems. Firstly, in experiments that require scanning of the $B$ field (e.g. cross-relaxation (CR) spectroscopy\,\cite{Hall2016,Wood2016,Wood2017,Simpson2017,Broadway2017}) the magnet must be physically moved which limits the precision of the field strength selection. Secondly, the magnetisation of permanent magnets is temperature dependent. Without adequate control, variations in ambient temperature will affect the stability of the qubit frequency especially over long measurement times\,\cite{Lillie2017}.  This instability can limit the resolution in nanoscale NMR spectroscopy\,\cite{Staudacher2013} and the efficiency of hyperpolarisation of external spins based on cross-relaxation\,\cite{Broadway2017}. In this work, we use the fact that the temperature dependence of the field strength in a magnet can itself be utilized for $B$-field scanning\,\cite{Mamin2017}, limiting the need for physical scanning. Thermal control of the magnet has the second benefit of facilitating the suppression of variations in qubit frequency caused by changes in magnetic field, improving the sensitivity, resolution, and applicability of constant frequency experiments by removing the concern of detuned control pulses. Experimentally, we demonstrate and quantify these stability improvements by probing the polarisation lifetime of a small ensemble of \C{13} spins in diamond.

\begin{figure*}
	\includegraphics[width=\textwidth]{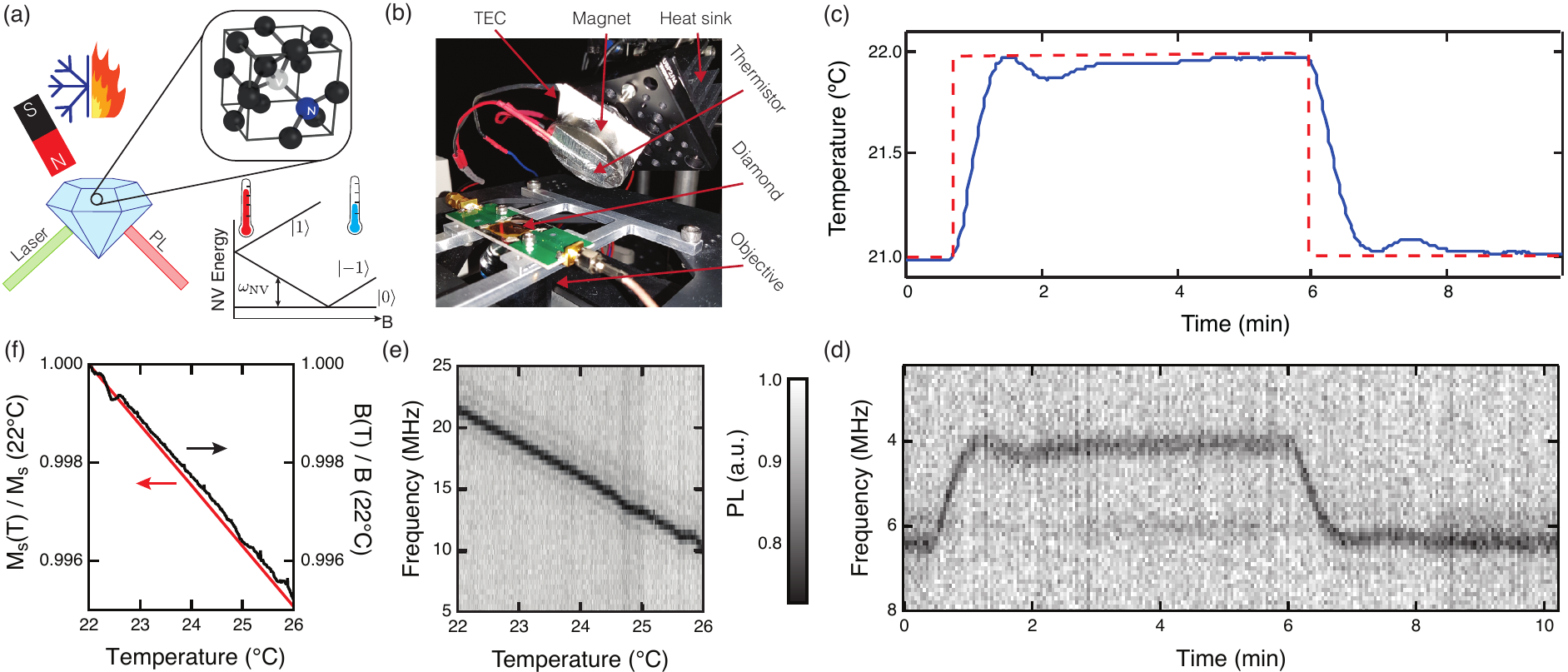}
	\caption{(a) Depiction of a nitrogen-vacancy (NV) centre in diamond excited with a green laser and subject to the field produced by a temperature controlled magnet. The NV state is read out via the red photoluminescence (PL). The NV \ket{\pm 1} states are split by the magnetic field applied along the quantization axis, where the magnet produces less field at hotter temperatures. (b) Picture of the magnet attached to a TEC module. (c) Step function in the set temperature (red dashed line) and the corresponding thermistor temperature reading (blue). (d) Corresponding optically detected magnetic resonance (ODMR) map from a single NV centre during the temperature pulse showing the \ket{0} $\rightarrow$ \ket{-1} spin transition with a background field $B \sim 1020$\,G. (e) ODMR mapped as a function of the set temperature. (f) Black curve: relative change in magnetic field $B$ vs temperature, as extracted from (e), using the relationship $\omega_NV=D-\gamma_e B$ gives a temperature coefficient of $\alpha_0 = -0.094$\%/\degC{}. Red curve: relative change in magnetisation M$_s$ vs temperature, as calculated from the specified coefficient of our neodymium magnet ($\alpha_0 = -0.12 \, \pm \, 0.1$\% \degC{}). Both plots are normalised by their respective values at $T = 22$\,\degC{}.}
	\label{Fig: intro}
\end{figure*}

%\section{Experimental set up}
 The frequency of the NV electron spin transition $\ket{0} \rightarrow \ket{-1} $(i.e. the qubit frequency) typically used in experiments is given by $\omega_{\rm NV} = D - \gamma_e B$ (Fig.\,\ref{Fig: intro}\,a), which has a zero field splitting of $D = 2.87$\,GHz with further splitting induced by a magnetic field along the quantization axis ($B$), with a defect gyromagnetic ratio $\gamma_e$ . Temperature control of the magnetisation of a permanent magnet is achieved by attaching a thermoelectric cooler (TEC) to the magnet as shown in Fig.\,\ref{Fig: intro}\,b. The control of the temperature was implemented through a feedback from a thermistor attached to the magnet on the opposite side to the TEC. This feedback was passed to a controller (TE technology TC-720) for proportional integral derivative (PID) control with Ziegler-Nichols method tuning. The magnet is a 50\,mm $\times$ 12.5\,mm  grade N38 neodymium disc with a NiCuNi coating. The specified temperature coefficient of the magnetisation $M_s$ around room temperature is $\alpha_0 = \frac{\partial M_s }{\partial t} =  -0.12\,\pm\,$0.1 \% /\degC{} with a maximum operating temperature between 80-240\,\degC{} and a Curie temperature of 310-370\,\degC{}. 
 
To test the ability to control the magnet temperature, we apply a 1\,\degC{} step function about ambient conditions (Fig. 1c, red dashed line) using the PID controller, and measure the thermal settling time of the magnet. Given the physical dimensions of the magnet, the thermal settling time for this modest temperature change (blue) is considerable (several minutes). For smaller temperature changes, in the regime $< 0.2$\,\degC{}, the settling time reduces to $< 1$ minute. The state dependent photoluminescence (PL) of the NV spin sub-levels\,\cite{Tetienne2012} allows the transition frequency between them to be measured by optically detected magnetic resonance (ODMR)\,\cite{Balasubramanian2008}. Performing an ODMR sweep every $10$\,s during the temperature change shows a direct correlation between the NV frequency  $\omega_{\rm NV}$ and the measured temperature of the permanent magnet, see Fig.\,\ref{Fig: intro}\,d. Here, the 1\,\degC{} change in temperature translates into a change of 2.5 MHz in $\omega_{\rm NV}$ at a field of $B \sim 1020$\,G (i.e. $\gamma_e B \sim 2860$\,MHz). We then mapped the response of the NV frequency to a ramp in temperature (Fig.\,\ref{Fig: intro}\,e), showing a linear response that is quantitatively consistent with the magnet specifications (Fig.\,\ref{Fig: intro}\,f). This response demonstrates a magnetic field scanning range of 0.5\% via varying the magnet temperature by only 4\,\degC{}  (from 22 to 26\,\degC{}), which corresponds to a 10\,MHz range at $B \sim 1020$\,G  (Fig.\,\ref{Fig: intro}\,e). With temperature variations up to 100\,\degC{} achievable with standard TECs, ranges as large as 12.5\% (i.e. $\sim 120$\,G in the magnetic range used in Fig.\,\ref{Fig: intro}\,e) may be achieved, which is sufficient to detect the spectra of electron-nuclear coupled systems with a large hyperfine splitting\,\cite{Wood2016}.  

%\section{Application to cross relaxation spectroscopy}

\begin{figure}
	\includegraphics[width=\columnwidth]{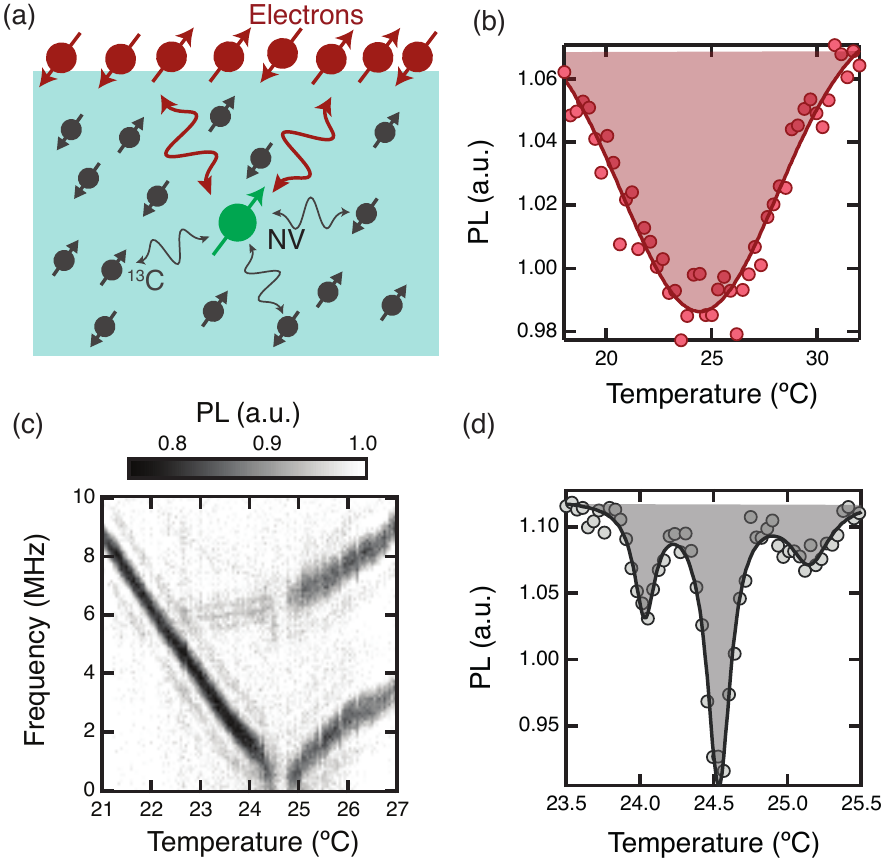}
	\caption{(a) Depiction of NV sensing external electrons (red) and internal \C{13} spins (dark grey). (b) CR spectrum ($\tau = 100 \mu$s) across the frequency range $\omega_{\rm NV} = 1426-1445$ MHz (via temperature scanning between  18 and 32\,\degC{}), showing a peak corresponding to unpaired electrons ($g=2$). (c) $^{14}$NV hyperfine structure across the GSLAC as mapped by temperature scanned ODMR. (d) CR spectrum ($\tau = 2 \mu$s) across the GSLAC ($\omega_{\rm NV} $ varied from -2 to 2 MHz) showing two peaks corresponding to \C{13} and a central peak due to the hyperfine induced level anticrossing. In (d), the measurement sequence alternates initialisation of the NV in \ket{0} and \ket{-1} before the free evolution time (see Fig. 4a(i)), to prevent net polarisation transfer to the \C{13} bath\,\cite{Broadway2017}. The data shown is the readout following initialisation in \ket{0} only.  }
	\label{Fig: sweeps}
\end{figure}

In previous works, frequency tuning was performed by a set of three motorised linear translation stages (PI M-511) allowing XYZ position control\,\cite{Wood2016}, or via simple manual stages for non-scanning experiments\,\cite{Lillie2017}. For CR spectroscopy it is desirable to implement small changes in frequency as this defines the maximum possible spectral resolution of the technique. We compare with our current mechanical control setup, with a 400\,nm axial step size; around the ground state level anti-crossing (GSLAC)\,\cite{Broadway2016} this step corresponds to a smallest NV frequency change of 65\,kHz. For the temperature control setup described here 0.01\,\degC{} steps in the magnet temperature were possible, corresponding to a change of $26\,$kHz. While the magnetisation range accessible by temperature control is not as large (physical motion allows for orders of magnitude variations in $B$) this option increases resolution by a factor of two, in addition to being significantly cheaper. Furthermore it can be implemented in a hybrid system to take advantage of both methods offering range and potentially comparable precision to piezo controlled stages. 

We now demonstrate the use of temperature enabled magnetic field scanning to perform CR spectroscopy on a single NV spin\,\cite{Mamin2017}. CR spectroscopy is a technique that monitors the relaxation time ($T_1$) of the NV spin as a function of $B$. This produces a spectrum of the surrounding spins, which are seen as decreases in the NV $T_1$ at each resonance  $\omega_{\rm NV} = \omega_{\rm env}$ where  $\omega_{\rm env}$ is the Larmor frequency of an environmental spin. We first demonstrate the technique targeting on electronic spins located on the surface of the diamond (Fig.\,\ref{Fig: sweeps}\,a, red). To this end, we use a near-surface NV centre ($\sim 5-10$\,nm deep\,\cite{Wood2017}), and a permanent magnet fixed such that $B \sim 512$\,G. The magnet temperature is then controlled to finely scan $B$ across the free electron resonance $\omega_{\rm NV} = \gamma_e B = 1435$\,MHz. Fig.\,\ref{Fig: sweeps}\,b shows the spectrum obtained by measuring the PL intensity $I_{\rm NV}$ following a free evolution time $\tau=100\mu$s, which maps changes in the $T_1$ time as $I_{\rm NV} \propto \exp(-\tau/T_1)$. The dip centred at a temperature $T = 24$\,\degC{} (corresponding to $\omega_{\rm NV} = 1435$\,MHz as measured by ODMR) is the signature of unpaired electron spins assumed to be associated with surface termination states or dangling bonds. By tuning the magnetic field close to the GSLAC (where $\omega_{\rm NV} < 5$\,MHz), it is also possible to detect nuclear spin species (Fig.\,\ref{Fig: sweeps}\,a, black). Fig.\,\ref{Fig: sweeps}\,c shows an ODMR map obtained from a deep NV centre ($\sim$1 $\mu$m from the surface) by scanning the temperature from 21 to 27\,\degC{} near $B=1024$\,G. The avoided crossing in the higher temperature range (i.e. lower magnetic field) is characteristic of the GSLAC structure for the $^{14}$NV centre\,\cite{Broadway2016}. The CR spectrum, measured using $\tau = 2\,\mu$s, reveals three features (Fig.\,\ref{Fig: sweeps}\,d). The central feature is caused by state mixing at the GSLAC and the two outer peaks are the signature of \C{13} nuclear spins in the diamond lattice, which come into resonance with the NV before and after the GSLAC\,\cite{Broadway2016,Broadway2017}.

%\section{Magnetic stabilisation}

\begin{figure}
	\includegraphics[width=\columnwidth]{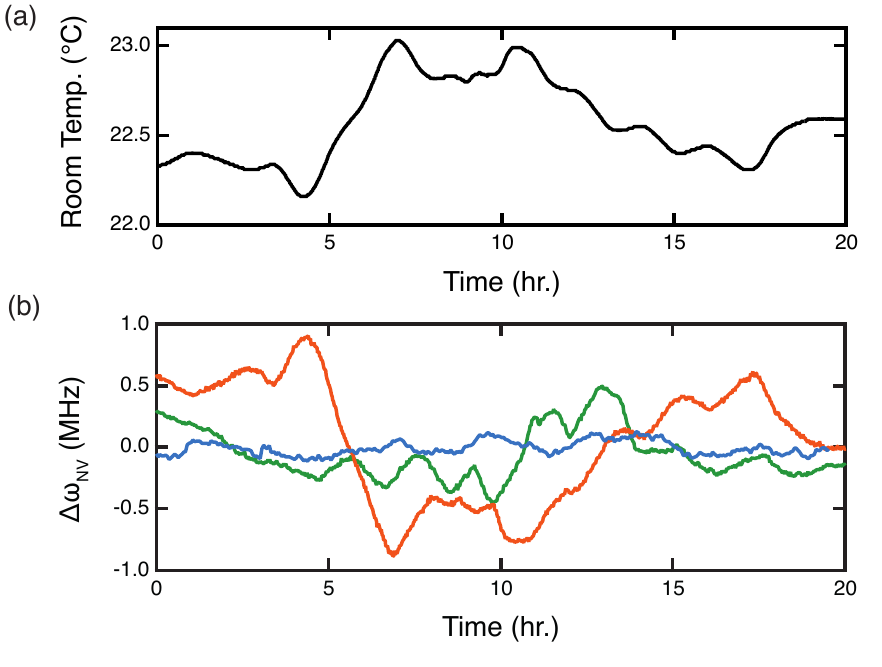}
	\caption{(a) Changes in the room temperature measured with the thermistor attached to the magnet in typical lab conditions, in the absence of active temperature control. (b) Variation in the NV frequency near the GSLAC, without temperature control (orange) which is strongly correlated with the temp change shown in (a), first order stabilisation (green) and second order stabilisation (blue). In the measurements leading to the green and blue curves, the room temperature fluctuations were similar to (a), see Supporting Information.}
	\label{Fig: stability}
\end{figure}

\begin{figure*}
	\includegraphics[width=\textwidth]{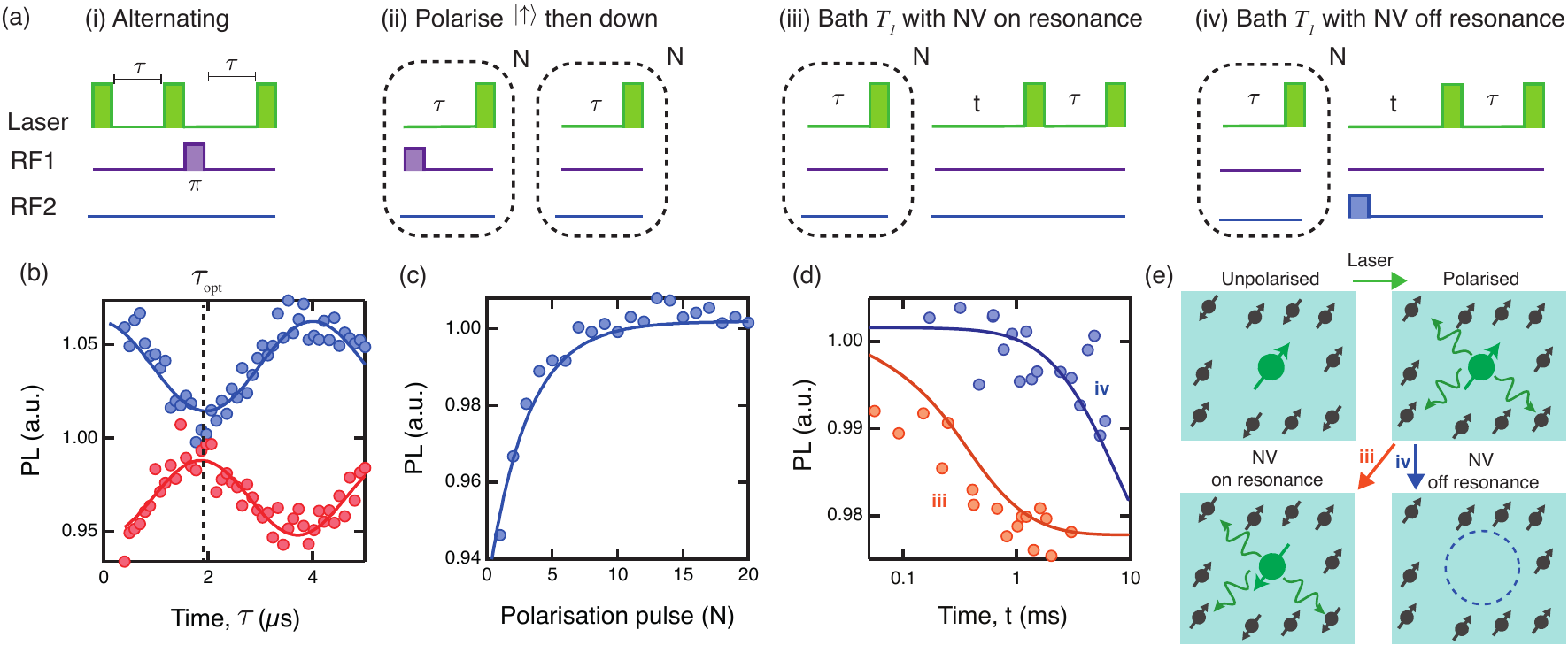}
	\caption{(a) Pulse sequences used for the study of the \C{13} polarisation where radio frequency RF1 is on resonance with \ket{0} $\rightarrow$ \ket{-1} transition (frequency $\approx 1.1$\,MHz) and RF2 is on resonance with \ket{0} $\rightarrow$ \ket{+1} (frequency $\approx 5.74$\,GHz). The pulse sequences have the effect of: (i) preventing polarisation of the bath, (ii) measuring the polarisation rate, (iii) measuring the lifetime of the spin bath polarisation with the NV on resonance and (iv) off resonance. (b) Interaction with \C{13} bath on resonance using the alternating sequence (i). (c) Dynamics of the \C{13} bath polarisation obtained using pulse sequence (ii) with $\tau =1.9 \mu s$. (d) Lifetime measurements of the polarisation of the \C{13}, showing a short lifetime $1/\Gamma_{pol} = 250(50) \mu$s when the NV spin was allowed to interact with the bath after polarisation (orange) and a significantly extended lifetime $1/\Gamma_{pol} \sim 10 $\,ms when the NV was flipped to an off-resonance state (blue). Solid lines numerical simulations of the spin bath under the pulse sequence (see Supporting Information). (e) Schematic of NV effect on bath polarisation.}
	\label{Fig: Bath T1}
\end{figure*}

A number of applications, such as high precision measurements\,\cite{Schmitt2017}, long decoupling sequences and hyperpolarisation of external spins \cite{Broadway2017,Fernandez-Acebal2017} require that the NV spin remain on resonance (with a driving field or a given spin species) for hours to days. While it is possible to post-process to remove off-resonance data, this is an undesirable solution for hyperpolarisation where off-resonant conditions limit the degree of polarisation transferred to the spin bath. By carefully controlling the temperature of the magnet, it is possible to significantly improve the long term stability of the NV frequency. In the absence of active temperature control, we observe variations in the room temperature (Fig.\,\ref{Fig: stability}\,a) that results in changes in the NV frequency on the order of 2 MHz over the course of 20 hours (Fig.\,\ref{Fig: stability}\,b, orange line). Including the PID loop to maintain a constant temperature of the magnet (as measured by the thermistor) reduces the change in frequency to 800\,kHz peak to peak (Fig.\,\ref{Fig: stability}\,b, green line), however, there is still an effect on the magnetisation from changes in the room temperature (see Supporting Information for details). This effect can be reduced by setting up a second order correction loop based on the measured ambient temperature, such that the changes to the magnet temperature set point are defined as
\begin{equation}
\Delta T_\text{M} = -\frac{\alpha_\text{R}}{\alpha_\text{M}} \Delta T_\text{R} + \frac{\alpha_\partial}{\alpha_\text{M}} \frac{\partial T_\text{R}}{\partial t}
\end{equation} 
where $\alpha_{\rm M} = 2.7$ MHz/\degC{} is the direct change in magnetisation from temperature, $ T_\text{M}$ and $ T_\text{R}$ are the thermistor reading of the magnet and room temperature, respectively, and $\Delta$ denotes difference from original value. $\alpha_\text{R}$ and $\alpha_\partial$ gives the change in magnetisation from the difference and derivative of the room temperature, respectively. This is simply a proportional derivative (PD) feedback loop acting on the set point,  which is combined with the PID loop acting on the TEC. Using this second order feedback reduces the changes to 200\,kHz (Fig.\,\ref{Fig: stability}\,b, blue line), offering an improvement in stability by an order of magnitude compared to a non-controlled temperature. This is a similar improvement to previous results using a temperature controlled stage with extensive insulation\,\cite{Haberle2017}. Importantly, the remaining fluctuations in the NV frequency are within the intrinsic ODMR linewidth for typical near-surface NV centres ($\sim 500$\,kHz), reducing the noise limit for long measurements to the intrinsic level. These fluctuations could be further reduced by simply changing to a common samarium-cobalt (SmCo) magnet which has a lower temperature coefficient $\alpha_0 = 0.04$. This would return a stability of $\sim 60$\,kHz and a resolution of $\sim 8$\,kHz, which is close to the smallest linewidths achievable \cite{Maurer2012}.

%\section{Polarisation investigation}

As a final experiment, we consider the polarisation of a small ensemble of \C{13} spins using cross relaxation induced polarisation (CRIP)\,\cite{Broadway2017}. By repeatedly initialising the NV in the same spin state and tuning its frequency to the \C{13} resonance, it is possible to gradually polarise the \C{13} bath, as demonstrated in Refs.\,\cite{Broadway2017,Wang2013}. Since the polarisation buildup occurs over long time scales (typically hours), instabilities in the NV frequency often limit the maximum level of polarisation achievable and the ability to monitor its evolution under a variety of conditions. Here, we use the improved stability to investigate the lifetime of the \C{13} polarisation. We first characterise the NV-bath interaction by using a sequence in which the NV spin is initialised in \ket{0} and \ket{-1} in alternance before the free evolution time $\tau$ (Fig.\,\ref{Fig: Bath T1}\,a(i)). By tuning the NV frequency on resonance with \C{13} (after the GSLAC, see Fig.\,\ref{Fig: stability}\,d) and varying $\tau$, we observe a coherent oscillation (Fig.\,\ref{Fig: Bath T1}\,b) whose frequency denotes the total coupling strength to the \C{13} bath\,\cite{Broadway2017}. Using an optimal evolution time $\tau_{\rm opt}=1.9\,\mu$s (half the period of the flip-flop oscillation), we then apply N polarisation steps to polarise the bath in the \ket{\uparrow} state (by initialising the NV in \ket{0}) followed by N steps to polarise the bath in \ket{\downarrow} (by initialising the NV in \ket{-1}), see sequence in Fig.\,\ref{Fig: Bath T1}\,a(ii). The resulting polarisation curve is plotted as a function of step number in Fig.\,\ref{Fig: Bath T1}\,c, and shows saturation of the polarisation after about 10 steps. Here only the spins within the coupling strength of $ \Gamma=1/\tau_{\rm opt}$ are probed, corresponding to a distance from the NV of $<1$\,nm. We also note that each polarisation step results in a maximum transfer of one quanta of polarisation and thus with N = 10 there is a maximum of 10 nuclear spins polarised.

The contrast between the polarised and unpolarised spin bath states allows measurements of the lifetime of the polarisation. To achieve this, the bath is initially polarised by applying $N=50$ polarisation steps, then after a time period $t$ the polarisation of the bath is measured again (Fig.\,\ref{Fig: Bath T1}\,a (iii)). By performing the measurement while still at the NV-\C{13} resonance, we obtain a decay of $T_1^{\rm pol} = 250 \pm 50\mu$s (Fig.\,\ref{Fig: Bath T1}\,d orange data). The decay time of this polarisation is limited by the NV spin $T_1$ (50$\pm$5 $\mu$s on resonance). While the inner core is polarised and no longer interacting with the NV spin, the outer shell which is unpolarised can interact with the NV during the probe time $t$, causing a drastically shortened NV $T_1$. Once the NV has decayed it acts as an additional noise source interacting with the polarisation of the spin bath (Fig.\,\ref{Fig: Bath T1}\,e bottom left). This effect can be suppressed by initialising the NV into the \ket{+1} state during the evolution time (Fig.\,\ref{Fig: Bath T1}\,a (iv)) effectively turning off the NV-\C{13} interaction (Fig.\,\ref{Fig: Bath T1}\,e bottom right). Including this NV spin flip extends the $T_1$ of the bath to the order of ten milliseconds (Fig.\,\ref{Fig: Bath T1}\,d, blue), which is limited by the phonon limited $T_1$ of the NV, on the order of milliseconds\,\cite{Jarmola2012}. These results agree well with theory, Fig.\,\ref{Fig: Bath T1}\,d solid lines, see Supporting Information for details. We note that the acquisition time for each curve in Fig.\,\ref{Fig: Bath T1}\,d was $\sim 20$ hours, during which the resonance condition was maintained; this would not have been possible to achieve without active temperature control with our apparatus.

In summary, a temperature control scheme was implemented on a permanent magnet and was shown to be capable of performing magnetic field sweeps to produce CR spectra of electron and nuclear spins. Controlling the magnet temperature improved the long term stability of the NV spin from $\pm$2\,MHz to $\pm$100\,kHz and could be further improved on with a SmCo magnet. With the increased stability, dynamics of the \C{13} polarised bath were explored. These results indicated that for small polarised regions the NV spin can significantly limit the lifetime of the polarisation imparted onto the bath. However, this effect can be mitigated by initialising the NV into a non-interacting state.

\textbf{Acknowledgements:} Original idea of temperature controlled magnetic scanning was brought to our attention by H. J. Mamin\,\cite{Mamin2017}. This work was supported in part by the Australian Research Council (ARC) under the Centre of Excellence scheme (project No. CE110001027). This work was performed in part at the Melbourne Centre for Nanofabrication (MCN) in the Victorian Node of the Australian National Fabrication Facility (ANFF). L.C.L.H. acknowledges the support of an ARC Laureate Fellowship (project No. FL130100119). J.-P.T acknowledges support from the ARC through the Discovery Early Career Researcher Award scheme (DE170100129) and the University of Melbourne through an Establishment Grant and an Early Career Researcher Grant. D.A.B and S.E.L are supported by an Australian Government Research Training Program Scholarship.

\bibliographystyle{apsrev4-1}
\bibliography{bib}

\section{Supporting Information}

\subsection{Magnetic stabilisation}

\begin{figure*}
	\includegraphics[width=\textwidth]{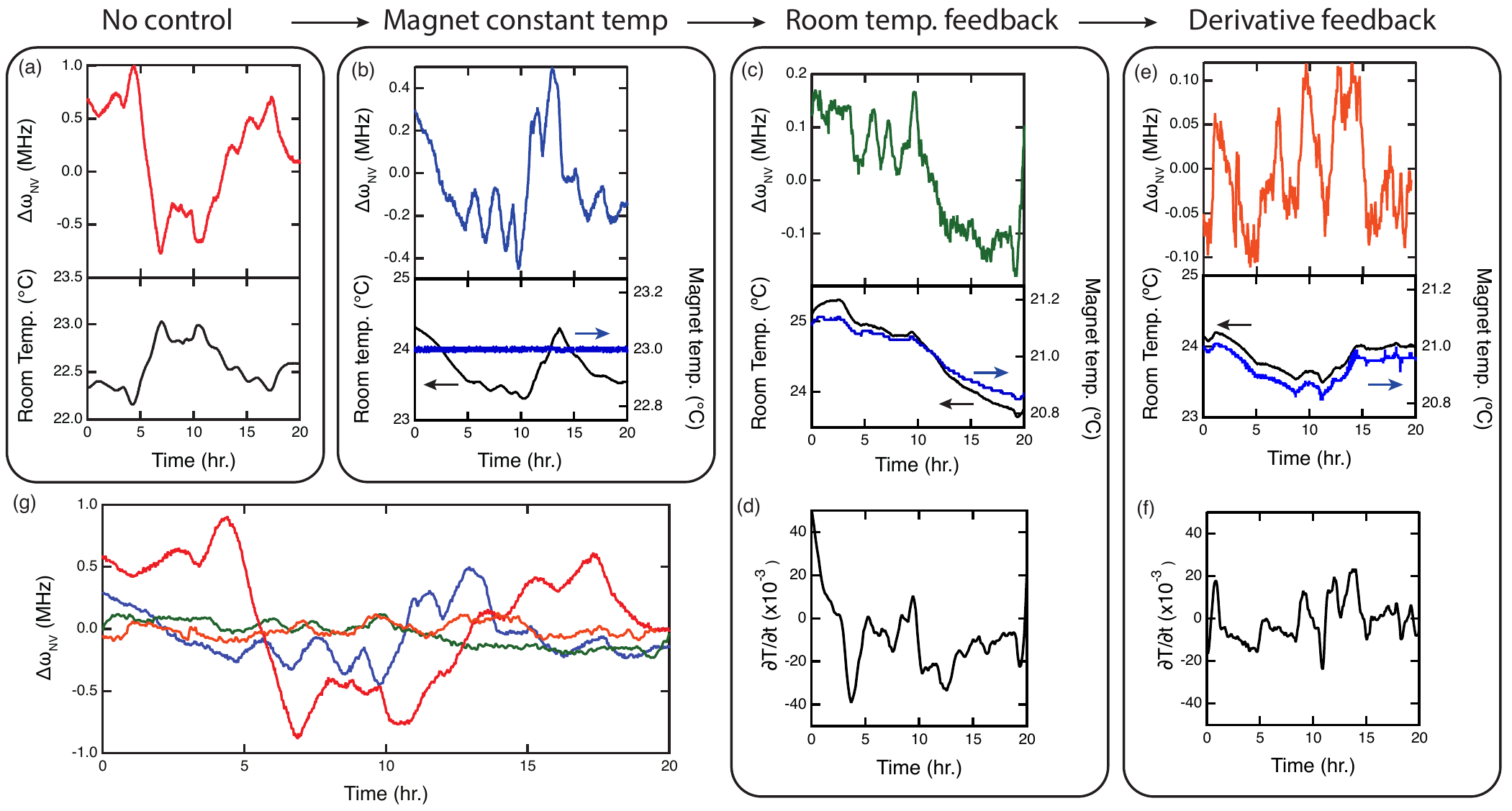}
	\caption{(a) Changes in the NV frequency over the period of 20 hours (top panel), and the room temperature over this period (bottom panel).  The temperature changes result in a variation in NV frequency of roughly 2 MHz peak to peak. (b) Changes in the NV frequency (top panel) while the TEC attempts to maintain a constant temperature of the magnet, reducing the fluctuation to 800 kHz. While the room temperature changed, the reading of the magnet temperature was constant (bottom panel). (c) Modulating the set temperature of the magnet based off the room temperature reduced the fluctuation to 400 kHz. (d) The derivative of the room temperature displayed in (c). (e) Inclusion of the derivative factor, further reducing the fluctuations to 200 kHz. (f) The derivative of the room temperature in (e). (g) Comparison of all the methods, colour matched to the other plots. }
	\label{Fig: SI stability}
\end{figure*}

Here we present additional details and measurements to illustrate the stabilisation of the magnet. Without stabilisation of the temperature there can be variations that results in changes in the NV frequency on the order of 2 MHz over the course of 20 hours, shown in Fig.\,\ref{Fig: SI stability}\,(a). These fluctuations in the NV spin transition frequency are directly correlated with the temperature fluctuations of the room shown in the bottom panel. Implementing the PID loop to maintain the magnet temperature reduces the fluctuations to 800 kHz, shown in Fig.\,\ref{Fig: SI stability}\,(b). While the thermistor reading is constant at the magnet surface there is still variation in the magnetisation of the magnet due to room temperature variations. 

Comparing Figs.\,\ref{Fig: SI stability}\,a and \ref{Fig: SI stability}\,b there is a difference in the direction of the change relative to the room temperature change, going from anti-correlated to correlated. When there is no control an increase in room temperature results in a decrease in magnetisation or NV frequency ($\Delta \omega_{\rm NV}$). However, with the PID loop on the changes are reversed, indicating that the change is due to the room temperature affecting the reading by the thermistor rather than changes to the temperature of the magnet. While extensive insulation would mitigate this issue it would also reduce the maximum achievable magnetic field at the NV, by increasing the minimum working distance. A thin layer of insulation was applied to the thermistor but showed no significant increase in stability. Instead we applied a feedback mechanism on the set temperature to compensate for this effect. Assuming that the change in magnetisation of the magnet is given by
\begin{equation}
\Delta M_s = \alpha_\text{M} T_\text{M} + \alpha_\text{R} T_\text{R}
\end{equation}
where $\alpha_{\rm M} = 2.7$ (MHz/\degC{}) obtained by measurements, $T_\text{M}$ and $T_\text{R}$ are the thermistor temperature reading of the magnet and the room temperature, and $\alpha_\text{R}$ is a constant that defines the effect of the room temperature on the magnetisation. Minimising the change gives,
\begin{equation}
T_\text{M} = -\frac{\alpha_\text{R}}{\alpha_\text{M}} T_\text{R}.
\end{equation}
Including this feedback factor reduces the magnetisation fluctuations from $800$ kHz to $400$ kHz, shown in Fig.\,\ref{Fig: SI stability}\,(c). To increase the stability further the relative speed of the fluctuation must be coincided. Comparing the derivative of the room temperature (Fig.\,\ref{Fig: SI stability}\,(d)) to the changes in magnetisation shows a clear correlation between some of the features. The set temperature can thus be changed to 
\begin{equation}
T_\text{M} = -\frac{\alpha_\text{R}}{\alpha_\text{M}} T_\text{R} + \frac{\alpha_\partial}{\alpha_\text{M}} \frac{\partial T_\text{R}}{\partial t}
\end{equation} 
where $\alpha_\partial$ is a constant that is modified to decrease the changes from the derivative. Including this feedback reduces the deviation in NV frequency to $200$ kHz (Fig.\,\ref{Fig: SI stability}\,(e)), which is within the line width of near surface NV centres. The derivative of the room temperature is shown in Fig.\,\ref{Fig: SI stability}\,(f). While some of the changes in $\Delta\omega_{\rm NV}$ are clearly seen in the derivative they are not completely removed. This is due to noise in the derivative being amplified and creating fast changes in $\alpha_\partial$. Therefore, there is a trade off between reducing the changes due to the derivative and introducing noise. While Savitzky-Golay filtering of the data improves this trade off it also introduces a lag time in-between the change in temperature and the reaction of the feedback. A comparison of all four control stages is shown in Fig.\,\ref{Fig: SI stability}\,(g), showing a significant improvement from the feedback system.

\subsection{Theory: Bath polarisation measurement}

\begin{figure*}
	\includegraphics{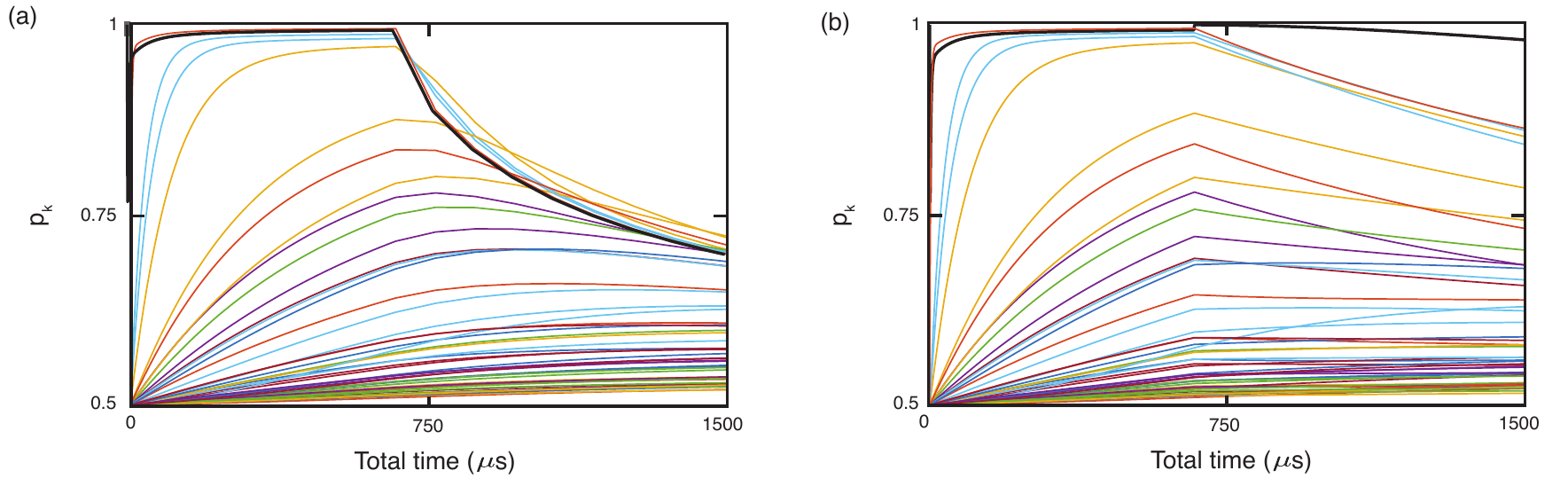}
	\caption{NV spin population (black) and population of 500 \C{13} spins for 50 polarisation pulses followed by an evolution time when the NV is left in \ket{0} (a) and for when the NV is put in the non-resonant \ket{+1} state (b).) }
	\label{Fig: SI theory}
\end{figure*}

Numerical simulations are performed on a random configuration of $N=$ 500-5000 spins at a density of 1.1\% \C{13}. The evolution of the N+1 system (where the NV is included) can be obtained using an evolution operator with a scattering matrix argument that is defined by the relevant hyperfine fields, such that
\begin{equation}
S = \text{Diag} \left[ \frac{1}{\pi} \frac{ \mathbf{A}_{sq} }{\sum_n \mathbf{A}_{sq}} -  \sum_n \sqrt{\mathbf{B}_{sq}}  \right] + \sqrt{\mathbf{B}_{sq}} - \frac{\mathbf{I}}{T_{1, E}}
\end{equation} 
where $\mathbf{A}_{sq} = (\mathbf{A}_{xx} - \mathbf{A}_{yy})^2 + 4\mathbf{A}_{xy}^2$ is the NV-\C{13} hyperfine matrix,  $\mathbf{B}_{sq} = (\mathbf{B}_{xx} - \mathbf{B}_{yy})^2 + 4\mathbf{B}_{xy}^2$ is the \C{13}-\C{13} hyperfine matrix, and $T_{1,E}$ contains both the boundary leakage of polarisation and rethermalisation of the \C{13} spins. This scattering matrix is effective at simulating both the polarisation build-up and the free evolution time. An example of a 50 polarisation pulses followed by a single evolution time is shown in Fig.\,\ref{Fig: SI theory} for both the NV in \ket{0} (a) and \ket{+1} (b). When the NV is in the \ket{0} state it decays and reduces the overall polarisation of the bath. However, when the NV is in the \ket{+1} state the polarisation of the bath remains relatively constant over the same time period.

\end{document}